\documentclass[3p,times]{elsarticle}

\usepackage{ecrc}

\usepackage{epstopdf}

\volume{00}

\firstpage{1}

\journalname{Nuclear Physics A}

\runauth{L. Apolin\'{a}rio et al.}

\jid{nupha}

\jnltitlelogo{Nuclear Physics A}

\usepackage{graphicx}
\usepackage{amsmath,amssymb}

\begin{document}

\begin{frontmatter}


%

\title{Energy loss and (de)coherence effects beyond eikonal approximation}


\author[label1,label2]{Liliana Apolin\'{a}rio}
\author[label1]{N\'{e}stor Armesto}
\author[label2,label3]{Guilherme Milhano}
\author[label1]{Carlos A. Salgado}

\address[label1]{Departamento de F\'{i}õsica de Part\'{i}culas and IGFAE, Universidade de Santiago de Compostela, \\15706 Santiago de Compostela, Galicia-Spain}
\address[label2]{CENTRA, Instituto Superior T\'{e}cnico, Universidade de Lisboa, \\Av. Rovisco Pais, P-1049-001 Lisboa, Portugal}
\address[label3]{Physics Department, Theory Unit, CERN, CH-1211 Gen\'{e}ve 23, Switzerland}

\begin{abstract}
The parton branching process is known to be modified in the presence of a medium. Colour decoherence processes are known to determine the process of energy loss when the density of the medium is large enough to break the correlations between partons emitted from the same parent. In order to improve existing calculations that consider eikonal trajectories for both the emitter and the hardest emitted parton, we provide in this work the calculation of all finite energy corrections for the gluon radiation off a quark in a QCD medium that exist in the small angle approximation and for static scattering centres. Using the path integral formalism, all particles are allowed to undergo Brownian motion in the transverse plane and the offspring is allowed to carry an arbitrary fraction of the initial energy. The result is a general expression that contains both coherence and decoherence regimes that are controlled by the density of the medium and by the amount of broadening that each parton acquires independently.
\end{abstract}


\end{frontmatter}



\section{Introduction}
\label{intro}

\par The use of large transverse momentum objects to probe the hot and dense medium formed in heavy-ion collisions has increased largely with RHIC and, more recently, with the LHC. The observation of this new state of matter is made indirectly through the modifications (generically called \textit{jet quenching}) of the hard probes when compared with proton-proton ($pp$) collisions. One of the possible observables concerns jets, a collection of particles that most likely are the result of the same parton branching. This process is well described in vacuum ($pp$ collisions) by perturbative QCD -- its development is controlled by vacuum splitting functions that determine the probability of a new splitting, and is mainly dominated by coherent emissions, where successive splittings follow angular ordering. In the presence of a medium, however, one should expect processes of energy loss, either by medium-induced gluon radiation or elastic scatterings, decoherence effects between successive emitters, and even the hadronisation, assumed to be universal for the $pp$ collisions, may change its pattern due to colour flow with the medium constituents.
\par There have been several theoretical developments that try to address independently radiative energy loss \cite{Bauer:2010cc,D'Eramo:2011zzb,Apolinario:2012vy}, broadening \cite{Idilbi:2008vm,D'Eramo:2010ak,Ovanesyan:2011kn,Ovanesyan:2011xy,Blaizot:2012fh} and decoherence effects \cite{MehtarTani:2010ma,MehtarTani:2011tz,CasalderreySolana:2011rz,Armesto:2011ir}. In this work, by calculating the medium-induced single gluon emission process beyond the eikonal approximation, we are able to unify the three previous effects in one single expression. For an extended version of this work we refer the reader to \cite{Apolinario:2014csa}.

\section{Amplitudes of the medium-induced single gluon radiation process}
\label{setup}

\par
\begin{figure}
\begin{center}
\includegraphics*[width=10.cm]{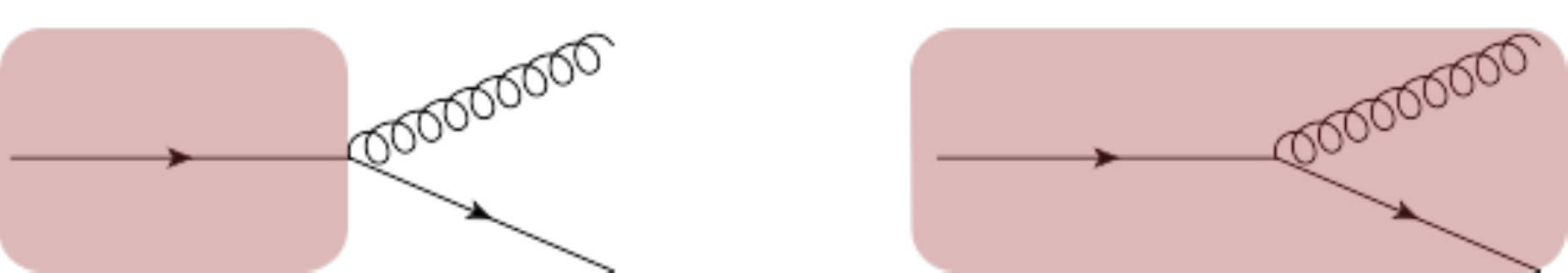}\\
\caption{
Schematic view of the process $q \rightarrow qg$ when only the initial quark interacts with the medium (left), that is represented as a shaded region, and when all particles propagate inside the medium (right).
}
\label{fig:diagrams}
\end{center}
\end{figure}

 In order to extended previous works to account for finite energy corrections to the energy loss, independent broadening of all propagating particles and colour correlation between different emitters, it is necessary to go beyond the soft limit, eikonal approximation and infinite medium. In that context, we consider the emission of a gluon with 4-momentum in light cone-coordinates $k = (\zeta p_{0+}, 0, \mathbf{k})$ off a quark with longitudinal momentum $p_{0+}$, originated from a hard process of amplitude $M_h(p_{0+})$, where $\zeta \in [0,1]$ and $\mathbf{k}$ is the transverse momentum of the gluon with respect to the quark. The final quark remains with $q = ((1-\zeta) p_{0+}, 0, \mathbf{q})$. Making use of the high-energy limit $(p_{0+} \gg |\mathbf{k}|, |\mathbf{q}|)$, the propagator that describes the random walk of a particle from time $x_{0+}$ with transverse coordinate $\mathbf{x_0}$ up to time $L_+$ at $\mathbf{x}$ through a frozen colour configuration of the medium is given by the Green's function:
\begin{equation}
\label{eq:G}
	G_{\alpha_f \alpha_i} (L_+, \mathbf{x}; x_{0+}, \mathbf{x_0} | p_+) = \int_{\mathbf{r}(x_{0+}) = \mathbf{x_0}}^{\mathbf{r}(L_+) = \mathbf{x}} \mathcal{D} \mathbf{r} (\xi) \exp \left\{ \frac{ip_+}{2} \int_{x_{0+}}^{L_+} d\xi \left( \frac{ d\mathbf{r}}{d\xi} \right)^2 \right\} W_{\alpha_f \alpha_i} (L_+, x_{0+}; \mathbf{r}(\xi)) \, ,
\end{equation}
\begin{equation}
\label{eq:W}
	W_{\alpha_f \alpha_i} (L_+, x_{0+}; \mathbf{x}) = \mathcal{P} \exp \left\{ ig \int_{x_{0+}}^{L_+} dx_+ A_- (x_+, \mathbf{x}) \right\} \, ,
\end{equation}
where eq. \eqref{eq:W} is the Wilson line that describes the colour rotation from $\alpha_i$ to $\alpha_f$ that the particle acquires due to the interactions with the medium. The contributions to the total amplitude will be given by:
\begin{equation}
\begin{split}
	\mathcal{T}_{out} = & \frac{-g}{4 (k\cdot q)} T^a_{BA_1} \int_{-\infty}^{+\infty} d\mathbf{x} \, d\mathbf{x}_{0}\,  \text{e}^{ -i \mathbf{x} \cdot (\mathbf{k} + \mathbf{q}) + i \mathbf{x}_0 \cdot \mathbf{p}_0} \,  G_{A_1 A} (L_+, \mathbf{x}; x_{0+}, \mathbf{x_0} | p_{0+}) \\
	& \times \bar{u} (q) \sl{\epsilon}_k^* (\sl{k} + \sl{q}) \gamma_+ \gamma_- M_h(p_{0+}) (2\pi) \delta(k + q - p_{0})_+ \, ,
\end{split}
\end{equation}
\begin{equation}
\begin{split}
	\mathcal{T}_{in} = & \frac{ig}{2} \int_{x_{0+}}^{L_+} dx_{1+} \int_{-\infty}^{+\infty} d\mathbf{x}_0 \, d\mathbf{x}_1 \, d\mathbf{y} \, d\mathbf{z} \, \text{e}^{ -i \mathbf{z} \cdot \mathbf{k} - i \mathbf{y} \cdot \mathbf{q} + i \mathbf{x}_0 \cdot \mathbf{p}_0 } G_{B B_1} (L_+, \mathbf{y}; x_{1+}, \mathbf{x_1} | q_+) T_{B_1 A_1}^{a_1} G_{A_1 A} (x_{1+}, \mathbf{x_1}; x_{0+}, \mathbf{x_0} | p_{0+}) \\
	& \times G_{aa_1} (L_+, \mathbf{z}; x_{1+}, \mathbf{x_1}| k_+) \bar{u}(q) \sl{\epsilon}_k^* \gamma_- M_h(p_{0+}) (2\pi) \delta (k + q - p_{0})_+ \, ,
\end{split}
\end{equation}
describing, respectively, the diagram in which only the initial quark propagates through the medium (figure \ref{fig:diagrams}, left) and the diagram in which all particles propagate through the medium (figure \ref{fig:diagrams}, right). $T_{A_f A_i}^{a}$ stands for the $SU(3)$ colour matrices. The differential spectrum will be given by:
\begin{equation}
\label{eq:spec}
	\frac{d^2 I_{tot}}{d\Omega_q d\Omega_k } = \frac{1}{\sigma_{el}} \left( \left\langle | \mathcal{T}_{out} |^2 \right\rangle + \left\langle | \mathcal{T}_{in} |^2 \right\rangle + \left\langle 2\, \text{Re}\left| \mathcal{T}_{in} \mathcal{T}_{out}^\dagger \right| \right\rangle \right)
\end{equation}
where the elastic cross-section is simply $\sigma_{el}  = \sqrt{2} (2\pi)^3 | M_h (p_{0+})|^2 $, the phase space $d\Omega_p = dp_+ d\mathbf{p} /(2 p_+ (2\pi)^3)$, $| \ldots |$ corresponds to the average over spins, polarisations and colours of the propagating particles and $\left\langle \ldots \right\rangle$ to the average over the ensemble of all possible medium colour configurations. In the limit of a non-existing medium (medium length or density $\rightarrow 0$), it is possible to recover the complete vacuum spectrum from the first term of eq. \eqref{eq:spec}.

\section{Medium averages}
\label{medium}

\par
\begin{figure}
\begin{center}
\includegraphics*[width=7.cm]{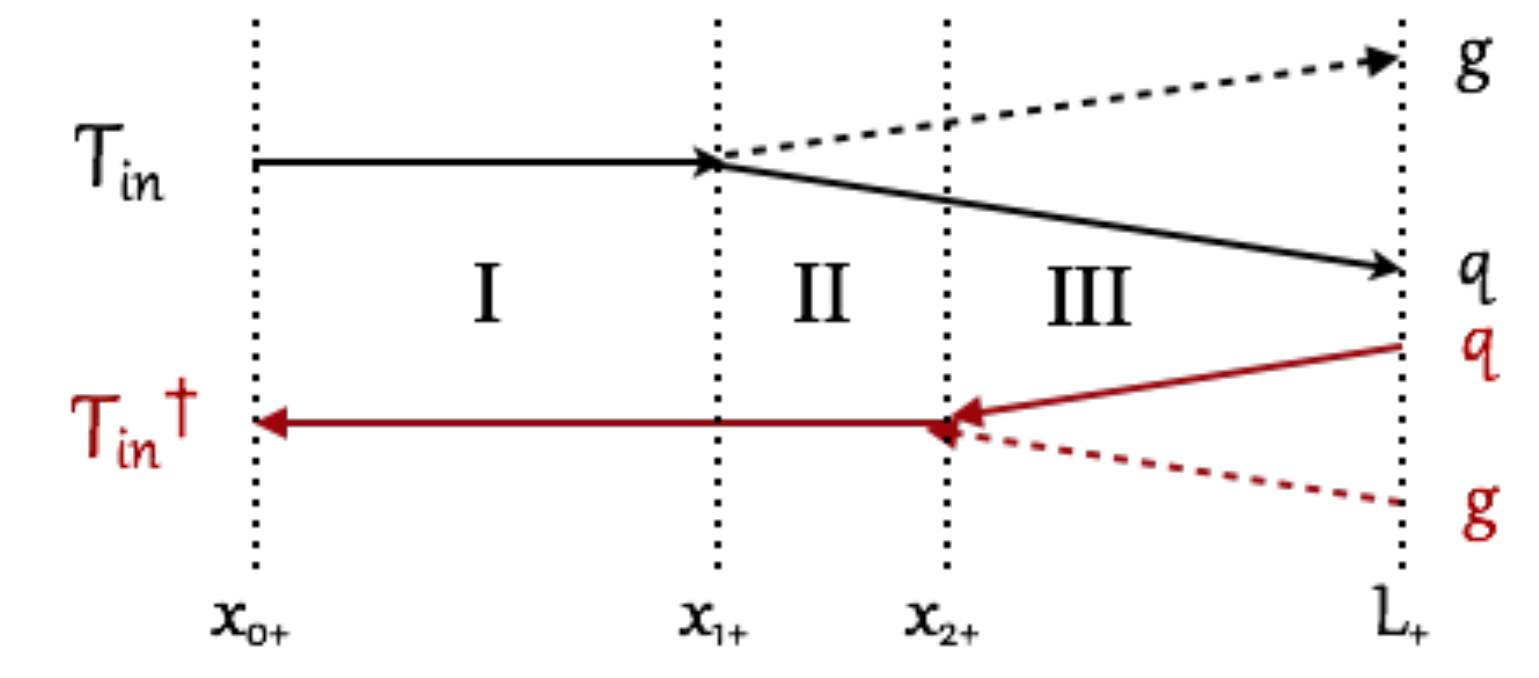}\\
\caption{
Schematic diagram of the $|\mathcal{T}_{in}|^2$ contribution, where upper arrows hold for the amplitude and lower ones for the conjugate amplitude. Solid lines represent the quarks and dashed lines the gluons. $x_{1+}$ and $x_{2+}$ are the emission points in the amplitude and conjugate one respectively.
}
\label{fig:diagramInIn}
\end{center}
\end{figure}

 To perform the medium averages, within the high-energy approximation, one can decompose the medium length into regions with a fixed number of propagators, as shown in figure \ref{fig:diagramInIn} for the $|\mathcal{T}_{in}|^2$ term. Factorising the colour structure from the transverse momentum dynamics (eq. \eqref{eq:G}), it is possible to evaluate each part independently. 

\par The colour structure implies the evaluation of several $n$-point correlation functions that can be computed systematically by making an infinitesimal expansion as:
\begin{equation}
	W_{ij} ( L_+,x_+; \mathbf{x}) = V_{i \alpha} (L_+,\tau; \mathbf{x}) \left[ \delta_{\alpha j} \left( 1 - \frac{C_F}{2} B(\tau, \mathbf{0}) \right) + i T^a_{\alpha j} A^a ( \tau, \mathbf{x} ) \right] \, ,
\end{equation}
where $V_{i\alpha}$ represents the Wilson line in the short trajectory $[\tau, L_+]$, and by considering only field correlators up to second order:
\begin{equation}
	\delta^{ab} B(x_+; \mathbf{x} - \mathbf{y}) = \left\langle A^a (x_+, \mathbf{x} ) A^b (x_+, \mathbf{y} ) \right\rangle \, .
\end{equation}
Applying this to a 2-point correlator, one can recover the well-known result as a function of the dipole cross-section, $\sigma$, medium density, $n(x_{+})$, and potential of each scattering centre, $a_-(\mathbf{q})$:
\begin{equation}
	\frac{1}{N} \text{Tr} \left\langle W(\mathbf{x}) W^\dagger(\mathbf{y}) \right\rangle = \text{e}^{- \frac{C_F}{2} v(\mathbf{x}-\mathbf{y})} = \exp \left\{ - \frac{C_F}{2} \int dx_+ \sigma(\mathbf{x} - \mathbf{y}) n(x_+) \right\} \, ,
\end{equation}
\begin{equation}
	\sigma (\mathbf{x}- \mathbf{y} ) = 2g^2 \int \frac{ d \mathbf{q} }{ (2\pi)^2Ê} | a_- (\mathbf{q}) |^2 \left(1 - \text{e}^{i \mathbf{q} \cdot (\mathbf{x} - \mathbf{y} ) } \right) \, .
\end{equation}

\par To evaluate the amount of broadening acquired by each particle, it is necessary to calculate the corresponding path integrals. For that, we use the dipole approximation \cite{Zakharov:1996fv,Zakharov:1998sv}: 
\begin{equation}
	n(\xi) \sigma (\mathbf{r}) \simeq \frac{1}{2} \hat{q} \mathbf{r}^2 + \mathcal{O} \left( \mathbf{r}^2 \ln \mathbf{r}^2 \right) \, ,
\end{equation}
where $\hat{q}$, the transport coefficient, characterises the typical squared transverse momentum that the particle acquires, per mean free path $\lambda$, and a semi-classical method \cite{FeynmanBook,Grosche:1998yu,Fujikawa}. 

\par Applying this method to figure \ref{fig:diagramInIn}, and translating colour representations to the fundamental one, region I assumes the calculation of two propagators, region II four propagators, and finally region III six propagators. At large $N_c$ the $6$-point function can be factorised into a dipole and an independent quadrupole that, within the approximations, reads:
\begin{equation}
\label{eq:quadruple}
	\text{Tr} \left\langle W^\dagger (\mathbf{x_g}) W (\mathbf{x_{\bar{g}}}) W^\dagger (\mathbf{x_{\bar{q}}}) W( \mathbf{x_q}) \right\rangle_{(x_{2+}, L_+)} = \text{e}^{N m_{22}} + \int_{x_{2+}}^{L_+} d\tau \text{e}^{N m_{11} (x_{2+}, \tau)} m_{12} (\tau) \, \text{e}^{N m_{22} (\tau, L_+) } \, ,
\end{equation}
where
\begin{equation}
\begin{split}
	m_{11} = - \frac{1}{2} \left[ v(\mathbf{x_{\bar{g}}} - \mathbf{x_{\bar{q}}}) + v (\mathbf{x_q} - \mathbf{x_g}) \right] \, , \ \ \ & \ \ \ 
	m_{22} = - \frac{1}{2} \left[ v(\mathbf{x_{\bar{g}}} - \mathbf{x_g}) + v (\mathbf{x_q} - \mathbf{x_{\bar{q}}}) \right] \, , \\
	m_{12} = - \frac{1}{2} \left[ v(\mathbf{x_{\bar{g}}} - \mathbf{x_{\bar{q}}}) + v (\mathbf{x_q} - \mathbf{x_g}) \right. & \left. - \, v(\mathbf{x_{\bar{g}}} - \mathbf{x_q}) - v (\mathbf{x_{\bar{q}}} - \mathbf{x_q}) \right] \, .
\end{split}
\end{equation}
The term $m_{11}$ represents a coherent propagation as it correlates the quark with the gluon, $m_{22}$ describes an independent propagation while $m_{12}$ is the swap term between the two propagation regimes. Therefore, the initial quark starts by propagating through the medium undergoing Brownian motion (region I). After the emission of the gluon, the two particles propagate coherently during the formation time, $\tau_{form} = x_{2+} - x_{1+}$, by definition (region II). Afterwards (region III), they can propagate completely decorrelated (first term of eq. \eqref{eq:quadruple}), or, alternatively (second term of eq. \eqref{eq:quadruple}), after the formation time the two particles continue colour correlated up to $\tau$ before undergoing an independent propagation. In the last case, the medium-induced gluon radiation is suppressed with respect to the former. Equation \eqref{eq:quadruple} can be re-written as a completely factorised piece, $\text{e}^{N m_{22} (x_{2+}, L_+)}$, times a decoherence parameters, $\Delta_{coh}$, that controls the suppression of the spectrum:
\begin{equation}
	\Delta_{coh} = 1 + \int_{x_{2+}}^{L_+} d\tau \text{e}^{N m_{11} (x_{2+}, \tau) - m_{22} (x_{2+}, \tau)} m_{12} (\tau) \, .
\end{equation}


\section{Conclusions}

\par In this work, by computing the single gluon emission spectrum off a quark beyond the eikonal approximation, we were able to associate Brownian motion to all propagating particles, extending, in this way, our previous results \cite{Apolinario:2012vy}. Moreover, our kinematic setup differs from \cite{Ovanesyan:2011kn} as we consider multiple soft interactions with the medium instead of a single hard scattering. Since we consider a finite medium, our results go beyond \cite{Blaizot:2012fh}, and in this case, a decoherence parameter $\Delta_{coh}$ is found. This parameter controls the factorisation of the spectrum of the two outgoing particles, taking into account the random walk of each. Therefore, it contains the same physics of colour decoherence as previous works in the antenna \cite{MehtarTani:2010ma,MehtarTani:2011tz,CasalderreySolana:2011rz,Armesto:2011ir}. 

\small{\textbf{Acknowledgments:} This work was supported by the European Research Council grant HotLHC ERC-2011-StG-279579, and by Funda\c{c}\~{a}o para a Ci\^{e}ncia e a Tecnologia of Portugal under projects  CERN/FP/123596/2011 and SFRH/BD/64543/2009}





\bibliographystyle{elsarticle-num}
\bibliography{Bibliography}



%
%
%

\end{document}